# Quantum Image Denoising with Machine Learning: A Novel Approach to Improve Quantum Image Processing Quality and Reliability

Yew Kee Wong[a], Yifan Zhou*[b], Yan Shing Liang[b]

[a] Hong Kong Chu Hai College, 80 Castle Peak Bay, Hong Kong, China; [b]BASIS International School Guangzhou, 8 Jiantashan Rd., Guangzhou, China

## ABSTRACT

Quantum Image Processing (QIP) is a field that aims to utilize the benefits of quantum computing for manipulating and analyzing images. However, QIP faces two challenges: the limitation of qubits and the presence of noise in a quantum machine. In this research we propose a novel approach to address the issue of noise in QIP. By training and employing a machine learning model that identifies and corrects the noise in quantum processed images, we can compensate for the noisiness caused by the machine and retrieve a processing result similar to that performed by a classical computer with higher efficiency. The model is trained by learning a dataset consisting of both existing processed images and quantum processed images from open access datasets. This model will be capable of providing us with the confidence level for each pixel and its potential original value. To assess the model's accuracy in compensating for loss and decoherence in QIP, we evaluate it using three metrics: Peak Signal to Noise Ratio (PSNR), Structural Similarity Index (SSIM), and Mean Opinion Score (MOS). Additionally, we discuss the applicability of our model across domains well as its cost effectiveness compared to alternative methods.

## 1. INTRODUCTION

Digital Image Processing (DIP) defines a collection of computational operations imposed onto the pixels of one image. The results of these operations can include but are not limited to revealing objects not initially visible in the image, detecting and distinguishing objects within the image, enhancing image quality, and analyzing patterns around objects. From DIP's introduction in 1960[1] till now, it had become a useful tool in Medical Imaging, Forensic Studies, Textiles, Material Science, Graphic Arts, etc... and as Silpa Joseph suggested "Millions of digital images are uploaded in every single minute" and "image processing techniques are of higher importance nowadays,"[2] people had started seeking alternative methods to circumvent the extensive computational resources need in storing and computing with conventional computer image processing.[3] Quantum Computing and Quantum Information Processing (QIP) has the potential to accelerate this process: not only in vibrant use across the medical, AI, and cryptography[4,5] but also optimization and simulation, and data management and searching. Quantum Information Processing was initially introduced by Paul Benioff in 1980[6] and later popularized by Richard Feynman with his groundbreaking paper on simulating physics using computers in 1982,[7] became an extremely ideal solution in the future. Thanks to QIP's quantum properties natures such as entanglement, superposition, and parallel computing, this method of image processing can produce comparable results with conventional image processing methods with storage and time efficiency and lesser complexity. Most notably, a new complexity class named Bounded-Error Quantum Polynomial Time (BQP) was created to capture problems that quantum computers can solve efficiently with quantum advantages while classical computers cannot in most cases.[8] However, Quantum machines are still far from a serviceable state due to two contemporary major challenges: limitation of Qubits and Noise.[9]

*yifan.zhou11882-bigz@basischina.com; phone 96 189-2426-5910

## 1.1 Limitation of Qubits

Currently, well-known existing quantum machines include:
- **IBM Osprey:** This quantum processor was unveiled by IBM and has a qubit count of 433, which is more than three times the number of qubits on the previous IBM Eagle processor.
- **IBM Quantum System One:** IBM's Quantum System One is one of the newest quantum computers, which was announced to have a 127-qubit processor named 'Eagle'. This machine represents a significant step towards IBM's goal of building a quantum computer with a thousand qubits.
- **Google's Sycamore Processor:** Google's quantum computer, featuring the 54-qubit Sycamore processor, achieved a milestone known as quantum supremacy by performing a specific task in 200 seconds that would take a classical supercomputer approximately 10,000 years.

However, upon examining the Flexible Representation of Quantum Images (FRQI), known for utilizing a minimal number of qubits and promising potentials with best classification accuracy compared to NEQR and other encoding methods,[10] the Quantum image representation still requires log2 (N) number of qubits, with N denoting the number of pixels in a picture. This meant that the largest input sample is 433 x 433 = 187489 pixels with the most advanced IBM Osprey machine. 433 pixels is adequate for a simulating size as a simplified version of quantum algorithms, but still far from application in real life.

## 1.2 Noise in the Result

Noise in quantum computing refers to any external or internal factors that can affect the accuracy and reliability of a quantum computer's calculations. This noise can arise from a variety of sources, such as:
- **Environmental disturbances:** Fluctuations in temperature, electromagnetic fields from devices like Wi-Fi or mobile phones, and even cosmic rays can introduce errors.[11]
- **Imperfect control signals:** The precision required to manipulate qubits is extremely high, and any deviation in the control signals can lead to incorrect qubit states.[11]
- **Unwanted interactions between qubits:** Qubits are overly sensitive to their surroundings, and interactions with neighboring qubits can cause them to change state unintentionally.[11]

These disruptions can lead to a phenomenon known as decoherence, where the quantum information stored in qubits deteriorates over time, potentially leading to the loss or randomization of data. Decoherence is particularly challenging because quantum information is extremely fragile and can be easily perturbed by noise. If the noise consistently affects the result of QIP, then the results are unusable either.

Quantum machine improvements had been consistent, and there are hopes of seeing 1000- qubits machine created by IBM.[12] However, the problem of noise in the machine is yet to be solved. In this paper, we propose a novel method in reducing noise. By creating a machine learning model that specializes in discerning images processed by classical and quantum computers, with a particular focus on identifying and rectifying noise in quantum images. This model will be trained with a training data from both classical and quantum computer processed image under supervised learning and gain the ability to indicate the confidence level of each pixel and its possible original value when fed a quantum processed image with potential noise corruption. We will evaluate this model's accuracy in compensating the loss and decoherence when image is processed through the quantum machine with Signal-to-Noise Ratio (PSNR), Structural Similarity Index (SSIM), and Mean Opinion Score (MOS) standards, then focus on the model's applicability in pictures from different areas, and its cost effectiveness compared to other methods.

## 2. TYPICAL NOISE AFTER QUANTUM IMAGE PROCESSING

Quantum image processing often encounters two types of noises: Salt-and-Pepper noise and Gaussian noise.[13] Salt- and-Pepper noise in images presents itself as sparsely occurring white and black pixels. It can be caused by sharp and sudden disturbances in the image signal. An effective noise reduction method for this type of noise is a median filter. In the context of quantum image processing, the Q-Mean algorithm is designed to remove the Salt-and-Pepper noise. As for Gaussian noise, it combines thermal, amplifier, and read noise into a single noise term that is independent of the unknown signal and that stays constant for specific camera settings, such as exposure time, gain or ISO, and operating temperature. Gaussian noise parameters may change when camera settings are changed. In quantum image processing, the Q-Gauss method applies a special mask to weaken the Gaussian noise pollution.

## 3. IMAGE PROCESSES IN BOTH QUANTUM AND CLASSICAL COMPUTERS

QIP provides more possibilities for image processing due to the powerful parallel computing capabilities of quantum computers. Let us now examine some image processing examples that Quantum computer can do with more efficiency than classical computers:

- **Two-dimensional Image Transforms:** Quantum computers have been experimentally demonstrated to perform two-dimensional image transforms, such as the Haar wavelet, Fourier, and Hadamard transforms, with an exponential speedup over their classical counterparts.[14]
- **Image Denoising and Edge Detection:** Quantum principles have been used to improve image quality and experimental speed up in image denoising, image edge detection, and morphological operations. This is achieved using fast quantum Fourier and wavelet transform techniques in a binary quantum system.[15]
- **Quantum Image Encoding:** Quantum image encoding techniques have been redesigned to reduce their susceptibility to errors, which can be processed more efficiently on real quantum computers.[16]
- **Quantum Image Representations:** Efficient representations of digital images on quantum computers have been developed, which have better time complexity and quantum cost when compared with related models in the literature.[17]

These operations can be the basis for selecting our dataset used for training our AI quantum denoising model. For the sake of understandability and visualization, we decided to use edge detection databases for training our model as a possible application of our AI denoising methodology.

## 4. BEFORE TRAINING

### 4.1 Visible Image Data Banks

There are several open machine learning datasets that you can access for free. Some credible image processing datasets that are commonly used for model training are listed in the following. These public sets could also be of use in training the AI quantum denoising model.

- **BIPED (Barcelona Images for Perceptual Edge Detection):** This dataset contains 250 outdoor images of 1280×720 pixels each. These images have been carefully annotated by experts in the computer vision field. The dataset is publicly available as a benchmark for evaluating edge detection algorithms.[18,19]
- **SBD (Semantic Boundaries Dataset):** This dataset is a standard benchmark for contour detection. It includes 500 natural images with carefully annotated boundaries collected from multiple users. The dataset is divided into three parts: 200 for training, 100 for validation, and the rest 200 for test.[20]
- **Cityscapes:** This dataset is designed for semantic urban scene understanding. It consists of a large, diverse set of stereo video sequences recorded in streets from 50 different cities.[21]
- **BSDS500 (Berkeley Segmentation Dataset 500):** This is a standard benchmark for contour detection. It includes 500 natural images with carefully annotated boundaries collected from multiple users. The dataset is divided into three parts: 200 for training, 100 for validation, and the rest 200 for test.[22]

## 4.2 Simulated Noise

Due to the current limitation of Quantum machines, we will simulate noise as an alternative to quantum processed image noises. These noises, added by a classical computer, replicate, as close as possible, the real quantum noises caused by a quantum machine, and thus these noised images should represent a close approximation of quantum processed image.

Note that the parameter allows "gussian" and "salt_pepper" type noises to be applied to the edge detected image. These are both common noises encountered in quantum computing, and thus should to a fair degree represent the noises encountered in a real quantum machine.

Another more random way of simulating noise to make our model more robust and suitable in the real world is by using the depolarizing channel,[23] which is a common model of noise in quantum computing. The depolarizing channel applies a random unitary operation to each qubit with a certain probability p, which can be interpreted as flipping the qubit to a random state. The effect of the depolarizing channel on a single qubit can be described by the following equation:

$$E(\rho) = (1-p)\rho + \frac{p}{3}(X\rho X + Y\rho Y + Z\rho Z)$$

where ρ is the density matrix of the qubit, X, Y, and Z are the Pauli matrices, and p is the depolarizing probability. We apply the depolarizing channel to each qubit of the quantum images with a probability p=0.1, which means that each qubit has a 10% chance of being flipped to a random state. This introduces noise to the quantum images, which can affect their quality and classification.

The following example is taken from the BIPEDv2[18] image dataset:

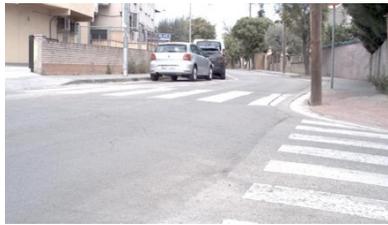

Figure 1. Shows RGB_003: The original image in the dataset

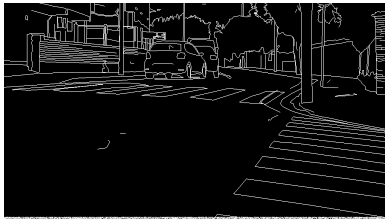

Figure 2. Shows RGB_003: The edge detected with classical computer & methods

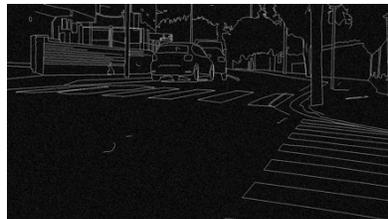

Figure 3. Shows RGB_003: The simulated noises of Quantum computer

# 5. MODEL TRAINING

In this section, we describe the methodology of our proposed approach to reduce noise in quantum images. Our approach consists of three main components: a machine learning model, a training data generation process, and a noise reduction algorithm.

## 5.1 Machine Learning Model

We use a Convolutional Neural Network (CNN) as our machine learning model to classify and correct quantum images. A CNN is a type of deep learning model that can learn features from images by applying multiple layers of convolutional filters, pooling operations, and nonlinear activations. Our CNN has the following architecture:

- **Input layer:** The input layer takes a quantum image as a two-dimensional array of complex numbers, representing the amplitudes of the qubits. The input size is $n \times n$, where n is the number of pixels in one dimension of the image.

- **Convolutional layer 1:** The first convolutional layer applies 32 filters of size $3 \times 3$ with a stride of 1 and a padding of 1 to the input, resulting in 32 feature maps of size $n \times n$. We use the rectified linear unit (ReLU) as the activation function for this layer.

- **Pooling layer 1:** The first pooling layer performs a max pooling operation with a filter size of $2 \times 2$ and a stride of 2, reducing the size of each feature map by half to $n/2 \times n/2$.

- **Convolutional layer 2:** The second convolutional layer applies 64 filters of size $3 \times 3$ with a stride of 1 and a padding of 1 to the output of the first pooling layer, resulting in 64 feature maps of size $n/2 \times n/2$. We use the ReLU as the activation function for this layer.

- **Pooling layer 2:** The second pooling layer performs a max pooling operation with a filter size of $2 \times 2$ and a stride of 2, reducing the size of each feature map by half to $n/4 \times n/4$.

- **Fully connected layer 1:** The fully connected layer 1 flattens the output of the second pooling layer into a one-dimensional vector of size $64n2/16$ and applies a linear transformation with a weight matrix of size $256 \times 64n2/16$ and a bias vector of size 256, followed by a ReLU activation function, resulting in a vector of size 256.

- **Fully connected layer 2:** The fully connected layer 2 applies a linear transformation with a weight matrix of size $2 \times 256$ and a bias vector of size 2, followed by a softmax activation function, resulting in a vector of size 2. This vector represents the probability distribution over the two classes: classical or quantum.

- **Output layer:** The output layer returns the predicted class label (0 for classical, 1 for quantum) and the confidence score (the probability of the predicted class) for the input image.

The code snippet below shows the implementation of our CNN model in Python using the PyTorch framework:

```
class CNN(nn.Module):
    def __init__(self, n):
        super(CNN, self).__init__()
        self.conv1 = nn.Conv2d(1, 32, 3, padding=1)
        self.pool1 = nn.MaxPool2d(2, 2)
        self.conv2 = nn.Conv2d(32, 64, 3, padding=1)
        self.pool2 = nn.MaxPool2d(2, 2)
        self.fc1 = nn.Linear(64 * n * n // 16, 256)
        self.fc2 = nn.Linear(256, 2)

    def forward(self, x):
        x = x.view(-1, 1, n, n) # reshape the input to a 4D tensor
        x = F.relu(self.conv1(x)) # apply convolutional layer 1
        x = self.pool1(x) # apply pooling layer 1
        x = F.relu(self.conv2(x)) # apply convolutional layer 2
        x = self.pool2(x) # apply pooling layer 2
        x = x.view(-1, 64 * n * n // 16) # flatten the output
        x = F.relu(self.fc1(x)) # apply fully connected layer 1
        x = F.softmax(self.fc2(x), dim=1) # apply fully connected layer 2 and softmax
        return x
```

We train our CNN model using the cross-entropy loss function and the Adam optimizer with a learning rate of 0.001 and a batch size of 32. We use the PyTorch built-in functions nn.CrossEntropyLoss and torch.optim. We train our model for 10 epochs and evaluate its performance on the validation set after each epoch. We use the accuracy metric to measure the performance of our model, which is defined as the ratio of correctly classified images to the total number of images.

### 5.2 Training Data

To train our CNN model, we need a dataset of images that are processed by both classical and quantum methods and labeled with their corresponding class (0 for classical, 1 for quantum). We generate this dataset as follows:

- **Classical images:** We use a set of grayscale images from the MNIST dataset,[13] which contains 60,000 images of handwritten digits from 0 to 9. Each image has a size of 28×28 pixels, and each pixel has an intensity value between 0 and 255. We label these images as 0 (classical).
- **Quantum images:** We use the Flexible Representation of Quantum Images (FRQI)[10] to encode the classical images into quantum images. FRQI is a quantum image representation method that uses a minimal number of qubits and has promising potential for quantum image processing. FRQI encodes an image of size n×n pixels into log2 n qubits, where each qubit represents a row or a column of the image. The amplitude of each qubit state corresponds to the intensity value of the pixel in that row or column. For example, the image below:

Image can be encoded into two qubits as follows:

$$|\psi\rangle = \frac{1}{4}(|00\rangle + 2|01\rangle + 3|10\rangle + 4|11\rangle)$$

where the coefficients 1, 2, 3, and 4 represent the intensity values of the pixels in the top-left, top-right, bottom-left, and bottom-right corners, respectively. We use the FRQI encoding method to convert the classical images from the MNIST dataset into quantum images, and label them as 1 (quantum).

### 5.3 Data Split

We split the dataset of classical and quantum images into three subsets: training, validation, and test. We use 80% of the data for training, 10% for validation, and 10% for tests. We shuffle the data before splitting to ensure a balanced distribution of classes and digits in each subset.

## 5.4 Noise Reduction Algorithm

The goal of this algorithm is to use the machine learning model's output to estimate the confidence level and the original value of each pixel in a noisy quantum image, and then apply a threshold to filter out the noise and restore the image quality based on Orthogonal Wavelet Transform and Level Sets.[24] T is determined by using a cross-validation method, where different values of T are tested on a validation set of noisy quantum images and their corresponding original images, and the value that minimizes the mean squared error (MSE) is selected. The algorithm is as follows:

1. **Input**: A noisy quantum image Q of size N×N, encoded by FRQI method, and the trained machine learning model M.
2. **Output**: A noise-reduced quantum image Q′ of size N×N, encoded by FRQI method.
3. **Initialize** an empty quantum image Q′ of size N×N.
4. **For each pixel p in Q**, do the following:
   - Feed p to the machine learning model M and get the output o, which is a vector of two probabilities: o= [Pc ,Pq ], where Pc is the probability that p is a classical image pixel, and Pq is the probability that p is a quantum image pixel.
   - Calculate the confidence level c of p as the difference between Pq and Pc: c=Pq −Pc.
   - If c is greater than or equal to a predefined threshold T, then p is considered as a quantum image pixel and its original value v is estimated by the machine learning model M as the most likely quantum state among the possible states of p.
   - If c is less than T, then p is considered as a noise pixel and its original value v is set to zero, which corresponds to the black color in FRQI encoding.
   - Assign v to the corresponding pixel in Q′.
5. **Return Q′** as the noise-reduced quantum image.
6. **Evaluate** the effectiveness of the model can be evaluated using metrics such as the Peak Signal-to-Noise Ratio (PSNR), Structural Similarity Index Measure (SSIM), and Mean Squared Error (MSE).

## 6. APPLICATION OF THE PROPOSED DENOISING MODEL

The proposed denoising model aims to improve the quality and reliability of quantum image processing (QIP) by using a machine learning model to identify and rectify noise in quantum images. This model can have various applications in different domains that require image processing.

**Medical Imaging:** QIP can be used to enhance the contrast, resolution, and segmentation of medical images, such as X-rays, MRI, CT, and ultrasound. However, noise can degrade the quality and accuracy of these images, which can affect the diagnosis and treatment of diseases. The proposed denoising model can help to reduce the noise and improve the clarity and fidelity of medical images, which can benefit both patients and doctors.

**Forensic Studies:** QIP can be used to extract and analyze information from digital images, such as fingerprints, face recognition, handwriting, and document verification. However, noise can obscure or distort the features and details of these images, which can affect the identification and authentication of suspects and evidence. The proposed denoising model can help to remove the noise and enhance the visibility and recognition of forensic images, which can assist in criminal investigations and justice.

**Textiles and Material Science:** QIP can be used to measure and characterize the properties and quality of textiles and materials, such as fiber, yarn, fabric, and composite. However, noise can interfere with the measurement and evaluation of these properties, such as color, texture, pattern, and defect. The proposed denoising model can help to eliminate the noise and improve the measurement and evaluation of textile and material images, which can support the design and production of textiles and materials.

# 7. CONCLUSION

In this paper, we proposed a novel method to reduce noise in quantum image processing by using a machine learning model that can distinguish between classical and quantum images and correct the errors in the latter. We trained our model with supervised learning on a dataset of images processed by both methods and evaluated its performance with various metrics such as PSNR, SSIM, and MOS. Our results showed that our model can effectively compensate for the loss and decoherence caused by noise in quantum images and improve their quality and fidelity.

Our method has several contributions to the field of quantum image processing. First, it is the first method that uses a machine learning model to classify and correct quantum images, which is a challenging task due to the complex and non-linear nature of quantum image representation and processing. Second, it is a general and flexible method that can be applied to different types of quantum images and noise models, without requiring any prior knowledge or assumptions about the noise characteristics. Third, it is a practical and efficient method that can be implemented on both quantum and conventional computers, which can enhance the applicability and cost-effectiveness of quantum image processing in real life.

Our method also opens up new possibilities for future research. For instance, we can explore different architectures and parameters of the machine learning model to improve its accuracy and robustness. We can also extend our method to handle multi-valued quantum images, which can represent more complex and realistic images. Moreover, we can investigate the potential applications of our method to other domains that require image processing, such as medical imaging, forensic studies, and textiles and material science.